\documentclass{mem}
\pdfoutput=1
\usepackage{natbib}\usepackage{txfonts}\usepackage{balance}
\usepackage{graphicx}
\usepackage[a4paper]{hyperref}
\usepackage{url}

\def\makeheadbox{
\parbox[t]{\textwidth}{%
  Proceedings 25th NSO Workshop \newline
      Eds.~A.~Tritschler, K.~Reardon, H.~Uitenbroek \newline
            Mem. S.A.It., in press}
	    \par\vspace*{-12ex}\mbox{}
}

\begin{document}

\title{On the detection of fast moving upflows in the quiet solar photosphere}

\subtitle{}

\author{
Th. \,Straus\inst{1} 
\and B. \,Fleck\inst{2}
\and S. M.\,Jefferies\inst{3}
\and M.\,Carlsson\inst{4}
\and T. D. \,Tarbell\inst{5}
}

\offprints{Th. Straus}

\institute{
INAF / Osservatorio Astronomico di Capodimonte, Via Moiariello 16, 80131 Napoli, Italy, \email{straus@oacn.inaf.it} 
\and
ESA Science Operations Department, c/o NASA Goddard Space Flight Center, Mailcode 671.1, Greenbelt, MD 20771, USA
\and
Institute for Astronomy, University of Hawaii, 34 Ohia Ku Street, Pukalani, HI 96768, USA
\and
Institute of Theoretical Astrophysics, University of Oslo, P.O. Box 1029 Blindern, N-0315 Oslo, Norway
\and
Lockheed Martin Solar and Astrophysics Laboratory, B/252, 3251 Hanover St., Palo Alto, CA 94304, USA
}

\authorrunning{Straus et al.}

\titlerunning{On the detection of fast moving upflows in the quiet solar photosphere}

\abstract{In our studies of the dynamics and energetics of the solar atmosphere, 
we have detected, in high-quality observations from Hinode SOT/NFI, 
ubiquitous small-scale upflows which move horizontally with supersonic 
velocities in the quiet Sun. We present the properties of these fast 
moving upflows (FMUs) and discuss different interpretations.
\keywords{MHD -- waves -- Sun: magnetic field -- Sun: oscillations -- Sun: chromosphere -- Sun: photosphere}
}
\maketitle

\section{Introduction}

Thanks to some remarkable advances in observational and theoretical
techniques, our understanding of the dynamics and energetics of the
lower solar atmosphere, in particular also of the quiet Sun magnetic
field, has significantly evolved in recent years.
\cite{2008ApJ...672.1237L} have reported the ubiquitous presence of
horizontal field in internetwork regions from observations with the
spectropolarimeter on Hinode. \cite{2008A&A...481L..25I} and
\cite{2009A&A...495..607I} have detected transient horizontal magnetic
field (THMF) to frequently emerge in both plage regions and the quiet
Sun, which might be energetically important.

On the theoretical side, 3-D numerical simulations have reached a
remarkable level of realism. The inclusion of radiative transfer
allows direct comparison of observational signatures derived from the simulations
with actual observations. Several groups have modeled the emergence of
horizontal flux (e.g.\,\cite{2007A&A...467..703C},
\cite{2008ApJ...687.1373C}, \cite{2008ApJ...680L..85S},
\cite{2008ApJ...679..871M}). The simulations of
\cite{2008ApJ...679..871M} span the remarkable height range from the
upper layers of the convection zone to the lower
corona. \cite{2002ApJ...564..508R} and \cite[][hereinafter B03]{2003ApJ...599..626B} 
have presented a detailed numerical model of the
propagation of magneto-acoustic-gravity (MAG) waves in magnetic
structures in the solar atmosphere. 

Here we report the detection of fast moving upflows (FMUs) in the
quiet solar photosphere. These approximately 1\arcsec\ small elements
appear in Dopplershift measurements, in the Mg~b$_2$ line obtained
with Hinode/NFI, to travel horizontally with supersonic velocities
(10--45~km~s$^{-1}$) in regions of low vertical magnetic field. Three
possible explanations are discussed below: (a) the signature of
emerging nearly horizontal flux ropes, (b) the signature of
propagating MAG waves along a horizontal flux tube appearing in
Dopplershift measurements in the Mg b$_2$ and Na D$_1$ lines, and c)
supersonic flows in flux tubes.

\section{Filtergram observations with SOT/NFI}

\begin{figure}
\center
\resizebox{0.85\hsize}{!}{
\includegraphics[clip=true]{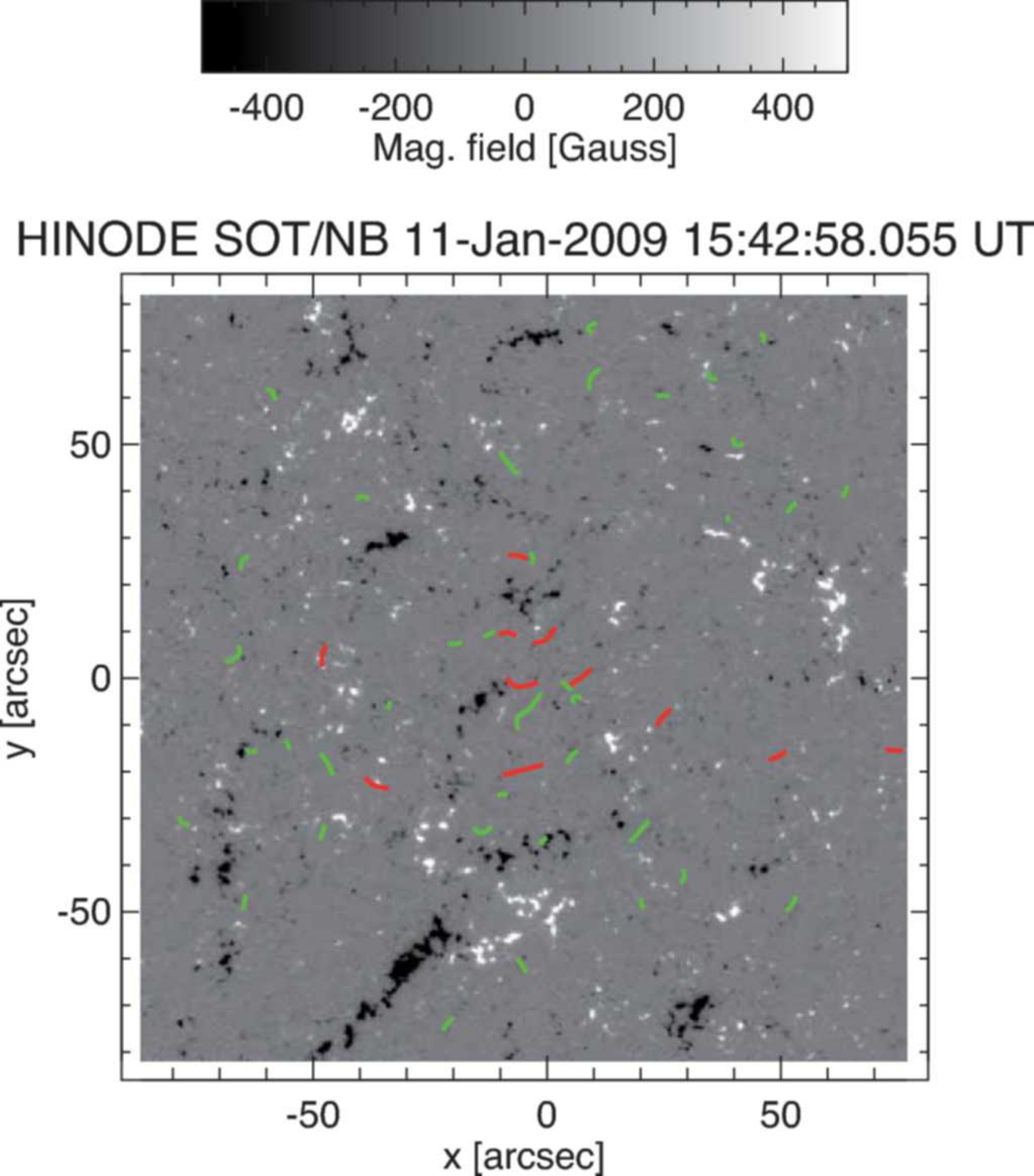}}
\caption{\footnotesize 
Summary of the supersonic events analyzed in this work. 
The green paths mark events found in the last 10 minutes 
of the time series and are therefore nearly simultaneous 
to the magnetic field map obtained after the run.
\label{Straus-fig:FMUmap}} 
\end{figure}

A 2-hour filtergram time-series has been obtained with SOT/NFI onboard
Hinode scanning 4 wavelength positions in the Mg b$_2$ line at
5173~m\AA\ on January 11th 2009, 13:28:34 -- 15:36:13~UT. A similar
time series has been obtained in the Na~D$_1$ line on January 7th. A
quick view of these latter data show similar features. For simplicity
we discuss only the Mg~b$_2$ data set in this short paper. The
observational setup has been designed to study wave propagation between
two different heights. The cadence achieved with this 4-wavelength
scan is 32~seconds. The offsets of the working points relative to the
nominal line center are $\pm 68$~m\AA\ (core) and $\pm 188$~m\AA\ (wing), respectively.

In order to calculate proxies of the velocities at two heights, the four
different time series have been spatially coaligned removing
instrumental jitter and interpolated to a common time frame using
spline interpolation. During the latter interpolation, missing or
defect data blocks have been interpolated. The red wing data set has
been used as spatial reference for image stabilization as these images
show the granulation pattern with highest contrast. During spatial
coalignment, the red wing data set has been stabilized using a cross
correlation algorithm comparing successive images. The blue wing and
red core data sets have then been aligned to this stabilized red wing
data set. Finally, the blue core data set has been aligned to the
stabilized red core data set.

After the coalignment, the velocity proxies have been calculated by
$S_v=(R-B)/(R+B)$ where $R$ and $B$ denotes the measured intensities
in the red and blue line positions, respectively, both near
the core (offset 68~m\AA) and in the wing (offset 188~m\AA). In
a realistic line profile the above defined signals are not perfectly
proportional to Dopplershift but can be used as a good proxy. A
detailed calibration to velocity is in progress for our study of the
vertical propagation characteristics of waves between the two
heights~\citep{StrausSP2}. In this work we use uncalibrated  proxy
values instead. We also calculate a proxy of the intensity fluctuation
as $S_I=(R+B)$, a calibration of which might be even more challenging
as this measure may be largely contaminated by line profile changes
due to velocities and magnetic flux. We can therefore use this last
proxy only as a coarse indication of the intensity fluctuations.

\begin{figure*}
\resizebox{0.9\hsize}{!}{
\includegraphics[clip=true]{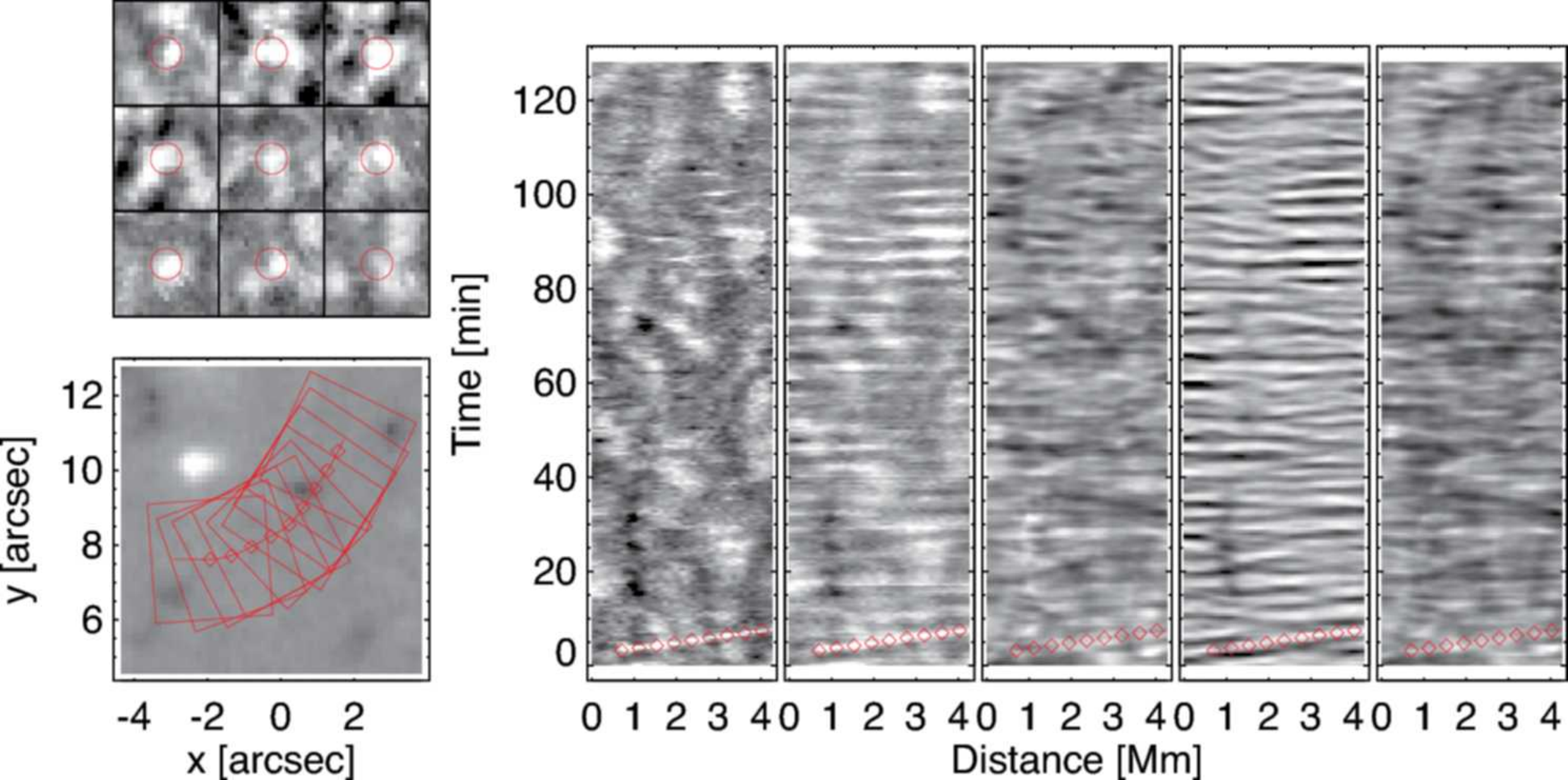}}
\caption{\footnotesize 
Details of one of the events displayed in Fig.~\ref{Straus-fig:FMUmap}: 
the lower left image shows the details of the event path over the magnetic 
field strength as in Fig.~\ref{Straus-fig:FMUmap}. The squares 
indicate the picture areas following the event as displayed in the 
upper left sequence of images. There, the order of sequence is from 
top left to bottom right, and individual frames with 32 seconds 
cadence are shown. The circles indicate an ideal structure of 
$0\farcs96$ diameter travelling at 13~km~s$^{-1}$ along the path. The 
direction of motion in these frames is always to the right. The 5 time-distance 
plots on the right side show (from left to right): wing velocity signal $S_v$ 
filtered as discussed in the text, unfiltered wing velocity signal, 
wing intensity signal $S_I$, core velocity signal, and core intensity 
signal. The symbols indicate the position of the ideal structure element 
mentioned before. 
\label{Straus-fig:Ex1}} 
\end{figure*}

\section{Results and discussion}

A first inspection of the time series of $S_v$ in the wing data set
reveals frequent events of FMUs all over the field-of-view. By
subtracting a spatially box-car averaged data set (with a spatial size
of the filter of 2\arcsec) these events become even more evident. A
movie of such a time series clearly reveals the ubiquitous presence of
the FMUs (see \href{http://www.oacn.inaf.it/~straus/public/67de45aa-9db7-4ae0-8abe-06394d56ebb2/Straus.mov}{online material}). As the later analysis will show, these
features appear to move horizontally with supersonic velocities up to
45~km~s$^{-1}$.

\begin{figure*}
\resizebox{\hsize}{!}{
\includegraphics[clip=true]{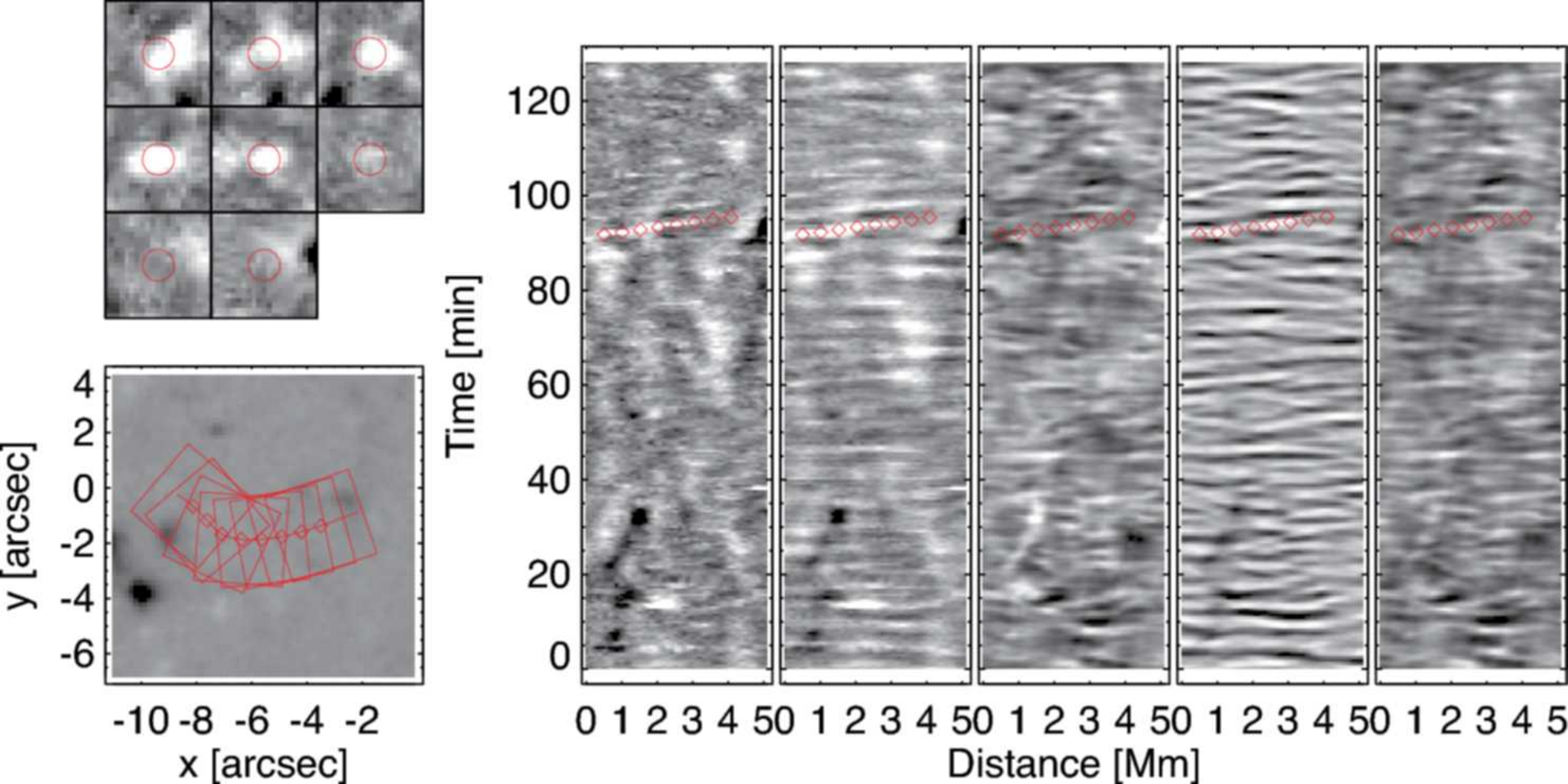}}
\caption{\footnotesize 
Details of a different event displayed in the same order as 
Fig.~\ref{Straus-fig:Ex1}. In this case, the ideal element was 
assumed to travel with 16~km~s$^{-1}$.
\label{Straus-fig:Ex2}} 
\end{figure*}

To characterize these features, we hand selected a number of events,
determined the propagation path and extracted time-distance diagrams
along the features' paths. Figure~\ref{Straus-fig:FMUmap} summarizes the
paths of the analysed events on top of a map of the line-of-sight
component of the magnetic field as measured by Hinode/SOT in the
Na~D$_1$ line 6 minutes after the time-series. This comparison is
useful mainly for characterizing the physical context of the fast
events, as we have access only to the line-of-sight component of the
field and only after the run. However, the curved paths of the events
we analyzed suggest a relation to magnetic fields
(e.g. Fig.~\ref{Straus-fig:Ex1} and \ref{Straus-fig:Ex2}). 

The FMUs have spatial scales of approximately 1\arcsec, can be
followed in our high-cadence time-series for up to 5~minutes, and tracing
paths as long as 4~Mm (see example in
Fig.~\ref{Straus-fig:Ex1}). They appear to be of rather impulsive
character, although weak evidence of a quasi-periodical behavior is
found in a few examples. They can be rather elongated in the direction
of propagation (see Fig.~\ref{Straus-fig:Ex2}). The shape of the FMUs
is often evolving during the motion. A higher temporal cadence would
be helpful to better identify their shape and evolution.

What are these FMUs? What drives them? We consider the following
three scenarios: (a) the upflow signature of plasma surrounding an
emergent THMF, (b) the manifestation of MAG waves propagating with
supersonic velocities along a horizontal flux tube, and (c) a
supersonic flow inside a flux tube.

In the context of an emerging THMF one could explain the FMUs as the
signature of the flows of the surrounding plasma driven at the point
of emergence, comparable to a horizontal rope emerging out of water
(see sketch in Fig.~\ref{Straus-fig:Scenario}). In this analogy the
water surface would represent the height where our velocity signal is
formed. The position of emergence at this surface will move
horizontally much faster than the emergence velocity if the 
rope is slightly inclined to the horizontal. The horizontal velocity
could then exceed the sound speed even if the flux rope emerges much
slower. The vertical velocity of the FMUs should then be approximately
the velocity with which the flux rope emerges. This scenario can
explain various properties of the observed FMUs. In fact, in this
scenario we would expect to see only upflows at the center of FMUs,
the spatial size of the features should be greater than the flux tube
section and could be rather elongated. The apparent curved paths of
the examples in Fig.~\ref{Straus-fig:Ex1} and \ref{Straus-fig:Ex2}
suggests curved flux tubes emerging. However, the geometry of such an
emergence is not completely clear. There seems to be little evidence
for the vertical "foot" points of these structures. One could also
expect to see two emergence points propagating away from each other if
the nearly horizontal structure is the central part of a
loop. Interestingly, we could not identify any such case so far.

\begin{figure}
\resizebox{\hsize}{!}{
\includegraphics[clip=true]{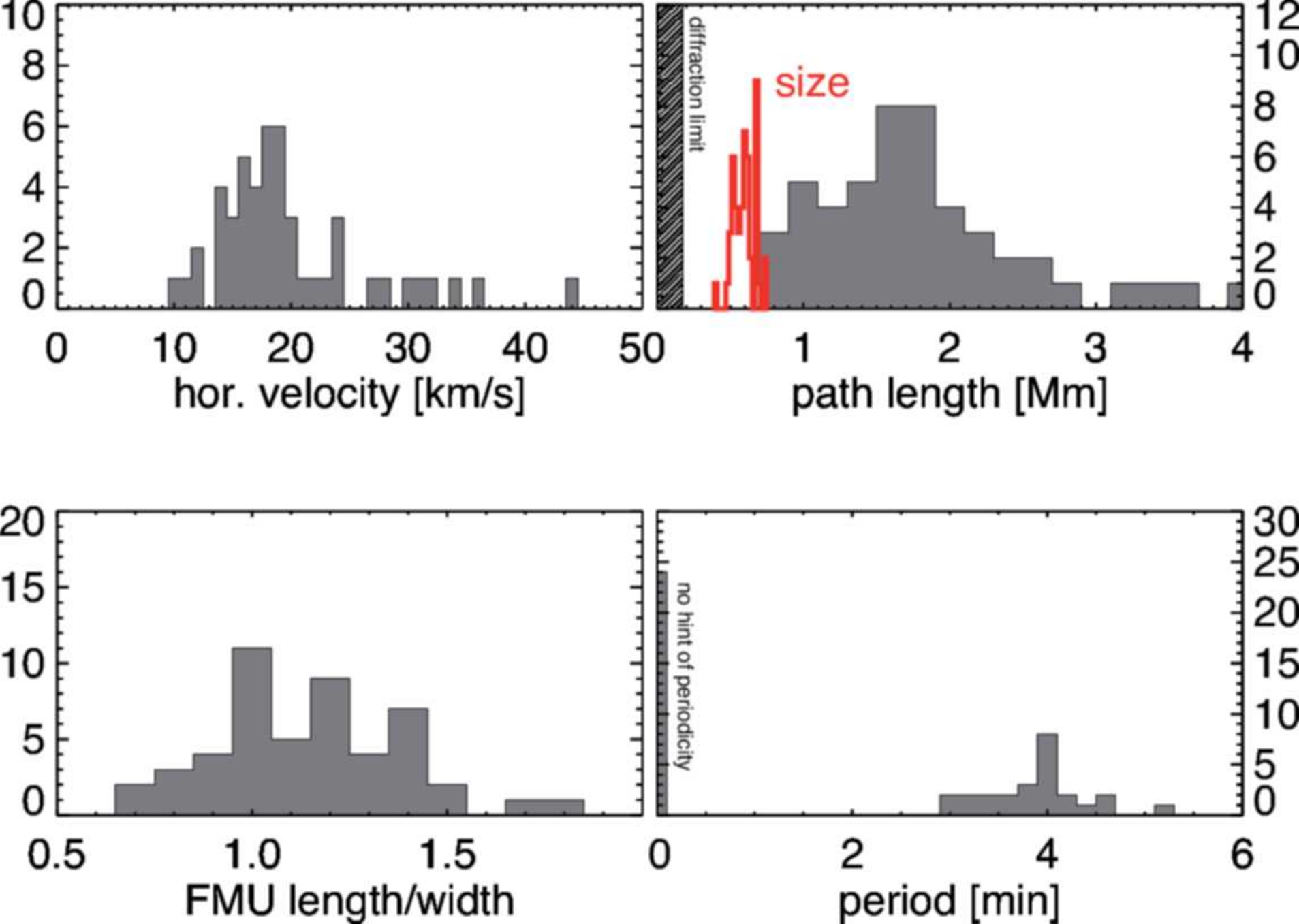}}
\caption{\footnotesize 
Histogram of properties of the 49 events shown in 
Fig.~\ref{Straus-fig:FMUmap}\label{Straus-fig:Hist}.} 
\end{figure}

\begin{figure}
\resizebox{\hsize}{!}{
\includegraphics[clip=true]{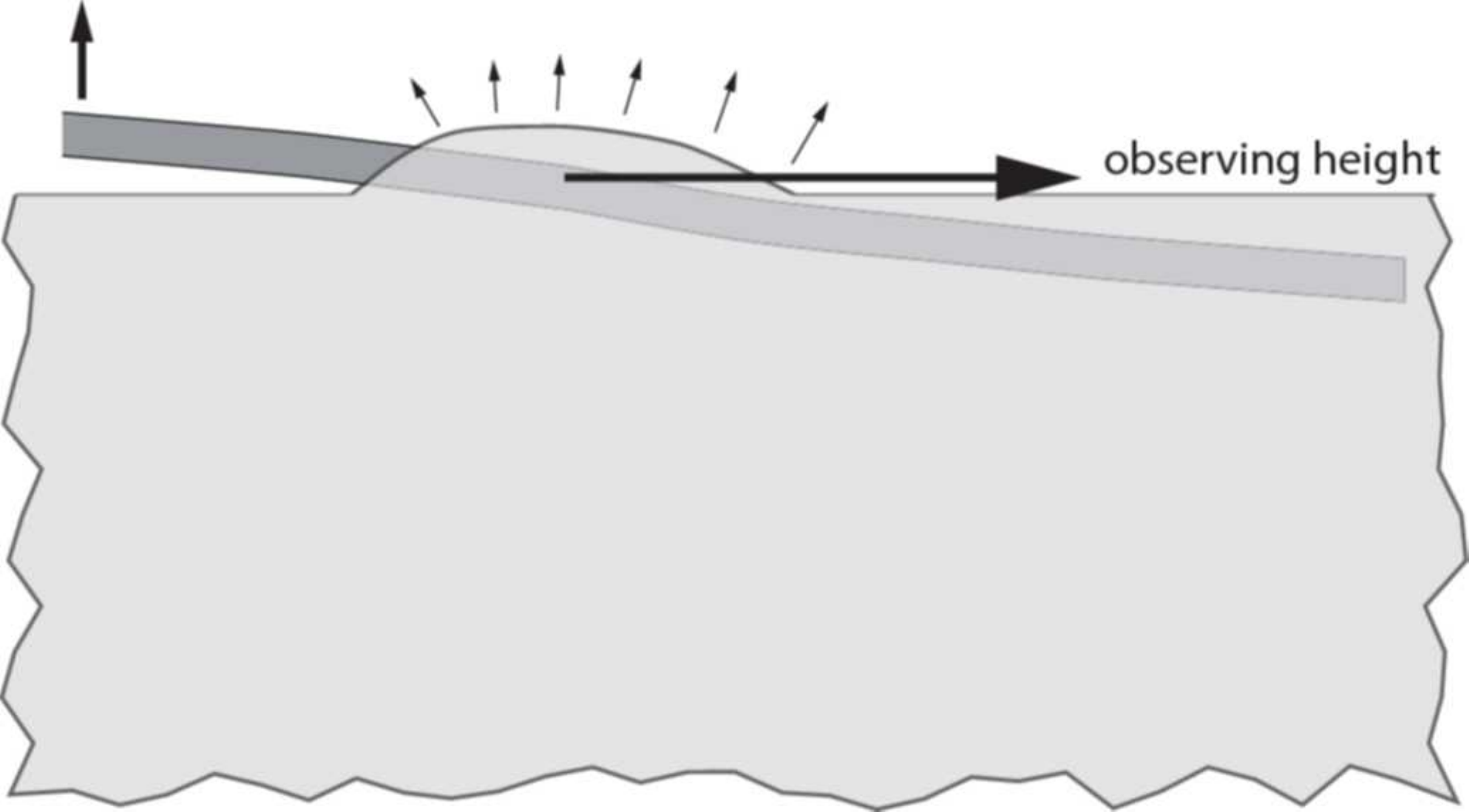}}
\caption{\footnotesize 
A schematic sketch of the scenario of an emerging 
horizontal flux tube that could explain the apparent 
super-sonic velocities.\label{Straus-fig:Scenario}} 
\end{figure}

On the other hand, a quasi-periodic behavior would strengthen the
hypothesis of a wave phenomenon along a horizontal flux tube. There is
no evidence for rather long coherence times like those of the MAG
waves studied by B03. As shown in the histograms of
Fig.~\ref{Straus-fig:Hist}, roughly half of the cases show some
evidence for wave trains of a couple of periods of the order of 4~minutes.

An interesting example for the following discussion of waves along or
flows inside a horizontal flux tube is shown in Fig.~2. This FMU is
moving horizontally at 13~km~s$^{-1}$ and can be followed along the clearly
curved path over a distance of nearly 4~Mm for nearly 5~minutes. In
the simplified picture of the wing signal being formed within a few
hundreds of kilometers in the low photosphere, this suggests a nearly
horizontal motion, otherwise the event would disappear quite rapidly
in the wing Doppler signal. It is interesting to recall here the
recent finding that nearly 80\% of the Sun's magnetic flux appears in
horizontal fields \citep{2008ApJ...672.1237L}. However, in the case of
an extremely cool cloud which makes the gas optically thick in the Mg
line, the signal would form essentially inside the cloud and could
therefore follow large height differences, though this hypothesis
seems to be incompatible with the results of our intensity proxy. 

Alternative interpretations of the FMUs are the propagation of a wave
along a static horizontal flux tube, or a flow inside such. To test
the compatibility of the various wave types with our observations we
searched for evidence of the events in the different observables (see
time-distance plots in Fig.~\ref{Straus-fig:Ex1} and
\ref{Straus-fig:Ex2}). The events are mainly visible as upward
velocity in the wing data set. Our filter removes a large part of the
p-mode signal in the time series and makes the events more
visible. However, the events are clearly visible also in the
unfiltered wing velocity. This suggests a transverse wave. On the
other hand, their signature in the intensity proxy is very weak, if
detectable at all. The expected velocity cross talk in $S_I$ makes any
spurious intensity signal  even more questionable. A Doppler signal of
the events is visible simultaneously in the higher-formed core signal
although the events are hard to isolate at that height, which is
dominated by the 3--5~minute oscillations. This could also be caused by
the broad contribution function of the core
signal~\citep{StrausSorrento} which has some overlap with the wing
signal.

B03 have shown that slow MAG waves in a vertical magnetic flux tube
can produce apparent horizontal propagation at supersonic speeds due
to a projection effect in case of a slightly inclined wave, even if
slow waves are limited to sonic speeds. However, to explain the event
of Fig.~\ref{Straus-fig:Ex1} that travels over nearly 4~Mm such an
explanation would require a plane wave extending over that distance in
a vertical flux element. Furthermore, the slow MAG waves should be
visible in intensity, for which we have no evidence. We therefore
exclude slow MAG waves as possible explanation for FMUs.

As for fast MAG waves, we would like to refer to the time-distance
plots of $\Delta\rho/\rho_0$ and $v_z/c_0$ in B03, who studied MAG
waves in a rather different geometry in a Sun-like atmosphere. These
diagrams indicate that fast waves should be visible in both vertical
velocity and intensity, except in the case of low-$\beta$. In the
latter case, only interference effects between different waves make a
weak signature in the intensity fluctuations. These effects, however,
are not relevant here, because of the impulsive character of the
FMUs. 

Flows inside flux tubes can reach real supersonic velocities. The
signature of the vertical velocity could then be interpreted as a
rising flux tube, or as the vertical component of the flow in an
inclined flux tube. However, it is unclear why the driving vertical
flows at the foot points of the flux tube are not visible. 

\section{Conclusions}

We report the detection of ubiquitous fast moving upflows (FMUs) in
the quiet Sun with SOT/NFI. These events show up in the Doppler signal
in the wing of the Mg~b$_2$ line. This signal is formed in the low
photosphere at about 150~km \citep{StrausSP2}. The small features of
approximately 1\arcsec\ size travel horizontally with apparent
supersonic velocities. Some of them show curved paths. The elements
can be followed up to 4~Mm and reach horizontal velocities of up to
45~km~s$^{-1}$. We discuss three mechanisms that might explain these
features: a) the emergence of nearly horizontal flux tubes, b) MAG
waves in horizontal magnetic fields, and c) supersonic flows in flux
tubes. The presently available observational material does not allow us
to distinguish between these options, and there might be other
explanations. For instance, one might speculate about a possible link
to the recently discovered "rapid blueshift events"
\citep{2009ApJ...705..272R}, which were identified as the on-disk
counterparts of type-II spicules \citep{2007PASJ...59S.655D}.

\begin{acknowledgements}
Th.S. acknowledges financial support by ESA and ASI. SJ acknowledges 
financial support from NSF. Hinode is a Japanese mission developed 
and launched by ISAS/JAXA, with NAOJ as domestic partner and NASA 
and STFC (UK) as international partners. It is operated by these 
agencies in co-operation with ESA and NSC (Norway). 
\end{acknowledgements}

\bibliographystyle{aa}

\end{document}